\documentclass[preprint,showpacs,preprintnumbers,amsmath,amssymb]{revtex4}
\usepackage{amssymb}
\usepackage{amsmath}
\usepackage{graphicx}
\usepackage{epsfig}
\usepackage{subfigure}
\usepackage{amsfonts}
\usepackage{CJK}

\begin{document}

\title{Deterministic generation of Greenberger-Horne-Zeilinger entangled states of cat-state qubits in circuit QED}

\author{Chui-Ping Yang$^{1}$}
\email{yangcp@hznu.edu.cn}
\author{Zhen-Fei Zheng$^{2}$}
\address{$^1$Department of Physics, Hangzhou Normal University, Hangzhou 310036, China}
\address{$^2$Key Laboratory of Quantum Information, University of Science and Technology of China, Heifei 230026, China}


\begin{abstract}
We present an efficient method to generate a Greenberger-Horne-Zeilinger
(GHZ) entangled state of three cat-state qubits (cqubits) via circuit QED.
The GHZ state is prepared with three microwave cavities
coupled to a superconducting transmon qutrit. Because the qutrit remains in
the ground state during the operation, decoherence caused by the
energy relaxation and dephasing of the qutrit is greatly suppressed. The GHZ
state is created deterministically because no measurement is involved.
Numerical simulations show that high-fidelity generation of a three-cqubit GHZ
state is feasible with present circuit QED technology. This proposal can be easily
extended to create a $N$-cqubit GHZ state ($N\geq 3$), with $N$ microwave or optical
cavities coupled to a natural or artificial three-level atom.
\end{abstract}
\pacs{03.67.Bg, 42.50.Dv, 85.25.Cp}\maketitle
\date{\today }

Cat-state qubits (cqubits), encoded with cat states, have drawn intensive
attention due to their enhanced life times \cite{s1}. Recently, there is an
increasing interest in quantum computing with cqubits. Schemes have been
presented for realizing single-cqubit gates and two-cqubit gates \cite
{s2,s3,s4}. Moreover, single-cqubit gates \cite{s5} and two-cqubit
entangled Bell states \cite{s6} have been experimentally demonstrated.
On the other hand, circuit QED, consisting of
microwave cavities and superconducting qubits or qutrits, has been
considered as one of the leading candidates for quantum information
processing (for reviews, see \cite{s7,s8,s9}).

The goal of this letter focuses on generation of Greenberger-Horne-Zeilinger (GHZ)
entangled states of cqubits via circuit QED. GHZ states have many
important applications in quantum information processing \cite{s10}, quantum
communication \cite{s11}, and high-precision spectroscopy \cite{s12}.
We will propose an efficient method to create three-cqubit GHZ
states, by using three microwave cavities coupled to a superconducting
transmon qutrit (a three-level artificial atom) [Fig.~1(a)].

This proposal has the following advantages. During the state
preparation, the qutrit stays in the ground state. Thus, decoherence from
the qutrit is greatly suppressed. The GHZ state is deterministically created
because this proposal does not require any measurement on the state of the coupler qutrit
or the state of the cqubits. Numerical simulations show that high-fidelity creation
of a three-cqubit GHZ state is feasible with current circuit QED technology. This
proposal can be easily extended to generate a $N$-cqubit GHZ state ($N\geq 3$),
with $N$ microwave or optical cavities coupled to a natural or artificial three-level atom.
To the best of our knowledge, this work is the first to demonstrate generation of GHZ entangled states with cqubits.

The three levels of the transmon qutrit are denoted as $|g\rangle $, $%
|e\rangle $ and $|f\rangle $ [Fig.~1(b)]. The $|g\rangle $ $\leftrightarrow $
$|f\rangle $ coupling for a transmon qutrit is forbidden or weak \cite{s13}.
Cavity $1$ is off-resonantly coupled to the $|g\rangle \leftrightarrow
|e\rangle $ transition of the qutrit but highly detuned (decoupled) from the
$|e\rangle \leftrightarrow |f\rangle $ transition of the qutrit. In
addition, cavity $l$ ($l=2,3$) is off-resonantly coupled to the $|e\rangle
\leftrightarrow |f\rangle $ transition of the qutrit but highly detuned
(decoupled) from the $|g\rangle \leftrightarrow |e\rangle $ transition of
the qutrit (Fig.~2). These conditions can in principle be satisfied by prior
adjustment of the qutrit's level spacings or the cavity frequency. For a
superconducting qutrit, the level spacings can be rapidly (within 1-3 ns)
adjusted \cite{s14,s15}. In addition, the frequency of a microwave
cavity can be quickly tuned in $1\sim 2$ ns \cite{s16,s17}.

\begin{figure}[tbp]
\centering
\includegraphics[bb=0 0 734 311, width=10.5 cm, clip]{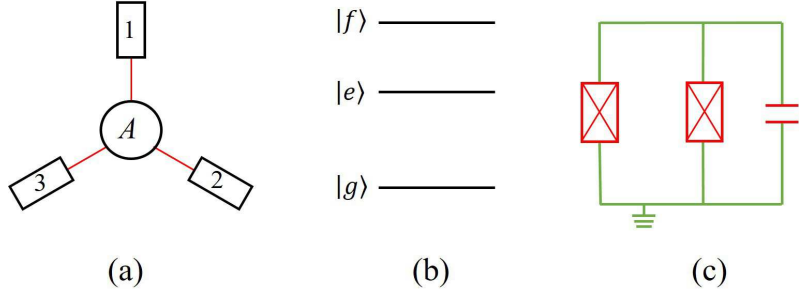}
\vspace*{-0.08in}
\caption{(Color online) (a) Diagram of three cavities coupled to a transmon
qutrit (labelled by $A$). Each square represents a cavity, which can be a
one- or three-dimensional cavity. The qutrit is capacitively or
inductively coupled to each cavity. (b) Level configuration of the transmon
qutrit, whose level spacing between the upper two levels is smaller
than that between the two lowest levels. (c) Circuit of a
transmon qutrit, which consists of two Josephson junctions and a capacitor.}
\label{fig:1}
\end{figure}

Under the above assumptions, the Hamiltonian, in the interaction picture and
after making the rotating-wave approximation, can be written as (in units of
$\hbar =1$)
\begin{align}
H_{\mathrm{I}}=g_{1}(e^{i\left\vert \delta _{1}\right\vert t}\hat{a}
_{1}^{+}\sigma _{eg}^{-}+h.c.)+\sum\limits_{l=2}^{3}g_{l}(e^{-i\left\vert
\delta _{l}\right\vert t}\hat{a}_{l}^{+}\sigma _{fe}^{-}+h.c.),
\end{align}
where $g_{1}$ and $g_{l}$ are the coupling constants, $\left\vert \delta
_{1}\right\vert =\omega _{c_{1}}-\omega _{eg},$ $\left\vert \delta
_{l}\right\vert =\omega _{fe}-\omega _{c_{l}},$ $\sigma _{eg}^{-}=|g\rangle
\langle e|$, $\sigma _{fe}^{-}=|e\rangle \langle f|$. The detunings $%
\left\vert \delta _{1}\right\vert $ and $\left\vert \delta _{l}\right\vert $
have a relationship $\left\vert \delta _{l}\right\vert =\left\vert \delta
_{1}\right\vert +\Delta _{1l},$ with $\Delta _{1l}=\omega _{fg}-\omega
_{c_{1}}-\omega _{c_{l}}>0$ (Fig. 2). Here, $\hat{a}_{1}^{+}$ ($\hat{a}%
_{l}^{+}$) is the photon creation operator of cavity $1$ ($l$), $\omega_{c_1}$ ($\omega
_{c_{l}}$) is the frequency of cavity $1$ ($l$ ($l=2,3$)); while $\omega _{fe},$ $%
\omega _{eg},$ and $\omega _{fg}$ are the $|e\rangle \leftrightarrow
|f\rangle ,$ $|g\rangle \leftrightarrow |e\rangle ,$ and $|g\rangle
\leftrightarrow |f\rangle $ transition frequencies of the qutrit,
respectively.

Under the large-detuning conditions $\left\vert \delta _{1}\right\vert \gg
g_{1}$ and $\left\vert \delta _{l}\right\vert \gg g_{l}$, the Hamiltonian
(1) becomes \cite{s18}
\begin{align}
H_{\mathrm{e},1}=& \lambda _{1}(\hat{a}_{1}^{+}\hat{a}_{1}|g\rangle \langle g|-%
\hat{a}_{1}\hat{a}_{1}^{+}|e\rangle \langle e|)  \nonumber \\
& -\sum\limits_{l=2}^{3}\lambda _{l}(\hat{a}_{l}^{+}\hat{a}_{l}|e\rangle
\langle e|-\hat{a}_{l}\hat{a}_{l}^{+}|f\rangle \langle f|)  \nonumber \\
& +\lambda _{23}\left( e^{i\bigtriangleup _{23}t}\hat{a}_{2}^{+}\hat{a}%
_{3}+h.c.\right) \left( |f\rangle \langle f|-|e\rangle \langle e|\right)
\nonumber \\
& +\sum\limits_{l=2}^{3}\lambda _{1l}(e^{-i\bigtriangleup _{1l}t}\hat{a}%
_{1}^{+}\hat{a}_{l}^{+}\sigma _{fg}^{-}+h.c.),
\end{align}%
where $\lambda _{1}=g_{1}^{2}/\left\vert \delta _{1}\right\vert $, $\lambda
_{l}=g_{l}^{2}/\left\vert \delta _{l}\right\vert $, $\lambda _{1l}=\left(
g_{1}g_{l}/2\right) (1/|\delta _{1}|+1/|\delta _{l}|)$, $\lambda
_{23}=\left( g_{2}g_{3}/2\right) (1/|\delta _{2}|+1/|\delta _{3}|),$ $%
\bigtriangleup _{23}=\left\vert \delta _{3}\right\vert -|\delta _{2}|$ $%
=\omega _{c_{2}}-\omega _{c_{3}},$ and $\sigma _{fg}^{-}=|g\rangle \langle f|
$. In Eq. (2), the terms in the first two lines describe the photon number
dependent stark shifts of the energy levels $|g\rangle $, $|e\rangle $ and $%
|f\rangle $; the terms in the third line describe the coupling between
cavities $2$ and 3; while the terms in the last line describe the $|f\rangle
$ $\leftrightarrow $ $|g\rangle $ coupling, caused due to the cooperation of
cavities $1$ and $l$ ($l=2,3$).

For $\bigtriangleup _{1l}\gg \{\lambda _{1},\lambda _{l},\lambda
_{23},\lambda _{1l}\}$, the Hamiltonian $H_{\mathrm{e},1}$ changes to \cite%
{s18}
\begin{align}
H_{\mathrm{e},2}=& \lambda _{1}(\hat{a}_{1}^{+}\hat{a}_{1}|g\rangle \langle g|-%
\hat{a}_{1}\hat{a}_{1}^{+}|e\rangle \langle e|)  \nonumber \\
& -\sum\limits_{l=2}^{3}\lambda _{l}(\hat{a}_{l}^{+}\hat{a}_{l}|e\rangle
\langle e|-\hat{a}_{l}\hat{a}_{l}^{+}|f\rangle \langle f|)  \nonumber \\
& +\lambda _{23}\left( e^{i\bigtriangleup _{23}t}\hat{a}_{2}^{+}\hat{a}%
_{3}+h.c.\right) \left( |f\rangle \langle f|-|e\rangle \langle e|\right)
\nonumber \\
& +\sum\limits_{l=2}^{3}\chi _{1l}(\hat{a}_{1}\hat{a}_{1}^{+}\hat{a}_{l}\hat{%
a}_{l}^{+}|f\rangle \langle f|-\hat{a}_{1}^{+}\hat{a}_{1}\hat{a}_{l}^{+}\hat{%
a}_{l}|g\rangle \langle g|),
\end{align}
where $\chi _{1l}=\lambda _{1l}^{2}/\Delta _{1l}$. When the levels $%
|e\rangle $ and $|f\rangle $ are initially not occupied, they will remain
unpopulated, because both $|g\rangle \rightarrow |e\rangle$
and $|g\rangle \rightarrow |f\rangle$ transitions are not induced by the Hamiltonian (3).
Thus, the Hamiltonian (3)
becomes $H_{\mathrm{e},3}=\lambda _{1}\hat{n}_{1}|g\rangle \langle
g|-\sum\limits_{l=2}^{n}\chi _{1l}\hat{n}_{1}\hat{n}_{l}|g\rangle \langle
g|, $ where $\hat{n}_{1}=\hat{a}_{1}^{+}\hat{a}_{1}$ ($\hat{n}_{l}=\hat{a}
_{l}^{+}\hat{a}_{l}$) is the photon number operators for cavities $1$ ($l$).

Suppose that the qutrit is initially in the ground state $\left\vert
g\right\rangle $. It will remain in this state throughout the interaction, as
the Hamiltonian $H_{\mathrm{e},3}$ cannot induce any transition for the
qutrit. In this case, the Hamiltonian $H_{\mathrm{e},3}$ is reduced to
\begin{equation}
H_{\mathrm{e}}=\lambda _{1}\hat{n}_{1}-\sum\limits_{l=2}^{3}\chi
_{1l}\hat{n}_{1}\hat{n}_{l},
\end{equation}%
which is the effective Hamiltonian governing the dynamics of the three
cavities.

\begin{figure}[tbp]
\centering
\includegraphics[bb=152 560 372 733, width=8.5 cm, clip]{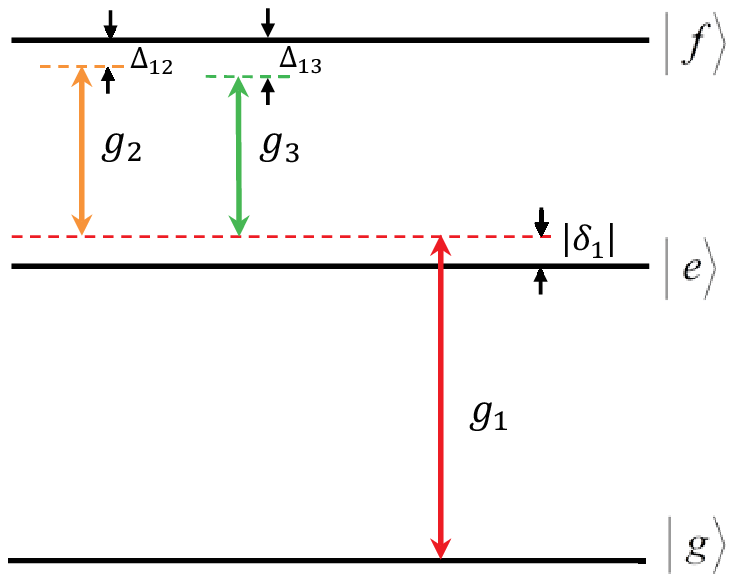}\vspace*{%
-0.08in}
\caption{(Color online) Cavity $1$ is far-off resonant with the $|g\rangle
\leftrightarrow |e\rangle $ transition with coupling strength $g_{1}$ and
detuning $\left\vert \protect\delta _{1}\right\vert $, while cavity $l$ ($%
l=2,3$) is far-off resonant with the $|e\rangle \leftrightarrow |f\rangle $
transition with coupling strength $g_{l}$ and detuning $\left\vert \protect%
\delta _{l}\right\vert $. Here, $\left\vert \protect\delta _{l}\right\vert $
($l=2,3$) is not drawn to simplify the figure. $\left\vert \protect\delta %
_{1}\right\vert =\protect\omega _{c_{1}}-\protect\omega _{eg}$, $\left\vert
\protect\delta _{l}\right\vert =\protect\omega _{fe}-\protect\omega %
_{c_{l}}=\left\vert \protect\delta _{1}\right\vert +\Delta _{1l},$ and $%
\Delta _{1l}=\protect\omega _{fg}-\protect\omega _{c_{1}}-\protect\omega %
_{c_{l}}>0$, with frequency $\protect\omega _{c_{1}}$ ($\protect\omega %
_{c_{l}}$) of cavity $1$ ($l$). The red line represents $\protect\omega %
_{c_{1}} $, while the brown and green lines represent $\protect\omega %
_{c_{2}} $ and $\protect\omega _{c_{3}}$, respectively.}
\label{fig:2}
\end{figure}

The unitary operator $U=e^{-i {H}_{\mathrm{e}}t}$ can be written as
$U=U_{1}\otimes \prod\limits_{l=2}^{3}U_{1l},$ where $U_{1}$ is a unitary
operator on cavity $1,$ while $U_{1l}$ is a unitary operator on cavities $1$
and $l,$ given by
\begin{equation}
U_{1}=\exp \left( -i\lambda _{1}\hat{n}_{1}t\right) ,~ U_{1l}=\exp \left(
i\chi _{1l}\hat{n}_{1}\hat{n}_{l}t\right) .
\end{equation}

The two logical states $|0\rangle $ and $|1\rangle $ of a $cqubit$ are encoded
with cat states of a cavity, i.e., $|0\rangle =N_{\alpha }^{+}(|\alpha
\rangle +|-\alpha \rangle )$ and $|1\rangle =N_{\alpha }^{-}(|\alpha \rangle
-|-\alpha \rangle ),$ where $N_{\alpha }^{\pm }$ are the normalization
coefficients. Because of $|\alpha \rangle =e^{-|\alpha
|^{2}/2}\sum\limits_{n=0}^{\infty }\frac{\alpha ^{n}}{\sqrt{n!}}|n\rangle $
and $|-\alpha \rangle =e^{-|\alpha |^{2}/2}\sum\limits_{n=0}^{\infty }\frac{%
(-\alpha )^{n}}{\sqrt{n!}}|n\rangle ,$ one has
\begin{equation}
|0\rangle =\sum\limits_{m=0}^{\infty }C_{2m}|2m\rangle ,\text{ }|1\rangle
=\sum\limits_{n=0}^{\infty }C_{2n+1}|2n+1\rangle ,
\end{equation}%
where $n$ and $m$ are non-negative integers, $C_{2m}=2N_{\alpha
}^{+}e^{-|\alpha |^{2}/2}\alpha ^{2m}/\sqrt{(2m)!},$ and $C_{2n+1}=$ $%
2N_{\alpha }^{-}e^{-|\alpha |^{2}/2}\alpha ^{2n+1}/\sqrt{(2n+1)!}$. Eq.~(6)
shows that the cat state $|0\rangle $ is orthogonal to the cat state $|1\rangle $,
independent of $\alpha $ (except for $\alpha =0$).

For $\chi _{1l}t=\pi ,$ the unitary operator $U_{1l}$ leads to the following
state transformation
\begin{align}
U_{1l}|2m\rangle _{1}|2m^{\prime }\rangle _{l}& =|2m\rangle _{1}|2m^{\prime
}\rangle _{l},  \nonumber \\
U_{1l}|2m\rangle _{1}|2n^{\prime }+1\rangle _{l}& =|2m\rangle
_{1}|2n^{\prime }+1\rangle _{l},  \nonumber \\
U_{1l}|2n+1\rangle _{1}|2m^{\prime }\rangle _{l}& =|2n+1\rangle
_{1}|2m^{\prime }\rangle _{l},  \nonumber \\
U_{1l}|2n+1\rangle _{1}|2n^{\prime }+1\rangle _{l}& =-|2n+1\rangle
_{1}|2n^{\prime }+1\rangle _{l},
\end{align}%
where we have applied $\exp \left[ i(2m)(2m^{\prime })\chi _{1l}t\right]
=\exp [i(2m)(2n^{\prime }+1)\chi _{1l}t]=\exp [i(2n+1)(2m^{\prime })\chi
_{1l}t]=1$ but $\exp [i(2n+1)(2n^{\prime }+1)\chi _{1l}t]=-1.$ By means of
Eq. (7) and according to Eq. (6), it is easy to find the following results
\begin{align}
U_{1l}|00\rangle _{1l}& =|00\rangle _{1l},\text{ }U_{1l}|01\rangle
_{1l}=|01\rangle _{1l},  \nonumber \\
U_{1l}|10\rangle _{1l}& =|10\rangle _{1l},\text{ }U_{1l}|11\rangle
_{1l}=-|11\rangle _{1l},
\end{align}%
which shows that a phase flip happens to the state $|1\rangle $ of cqubit $l$
when cqubit $1$ is in the state $\left\vert 1\right\rangle $.

The unitary operator $U_{1}$ leads to
\begin{eqnarray}
U_{1}|0\rangle _{1} &=&\sum\limits_{m=0}^{\infty }\exp \left[ i(2m)\lambda
_{1}t\right] C_{2m}|2m\rangle _{1},  \nonumber \\
U_{1}|1\rangle _{1} &=&\sum\limits_{n=0}^{\infty }\exp [i(2n+1)\lambda
_{1}t]C_{2n+1}|2n+1\rangle _{1}.
\end{eqnarray}%
For $\lambda _{1}t=2\pi ,$ we have $\exp \left[ i(2m)\lambda _{1}t\right]
=\exp [i(2n+1)\lambda _{1}t]=1.$ Hence, Eq. (9) becomes
\begin{equation}
U_{1}|0\rangle _{1}=|0\rangle _{1},\text{ }U_{1}|1\rangle _{1}=|1\rangle
_{1}.
\end{equation}

Now assume that the three cqubits are initially in the state $|\psi \rangle _{%
\mathrm{cq}}=\prod\limits_{l=1}^{3}\frac{1}{\sqrt{2}}\left( \left\vert
0\right\rangle _{l}+\left\vert 1\right\rangle _{l}\right) $, which can be
prepared from the state $\prod\limits_{l=1}^{3}\left\vert 0\right\rangle
_{l},$ by applying a driving pulse to cavity $l$ to obtain the state
rotation $\left\vert 0\right\rangle _{l}\rightarrow \frac{1}{\sqrt{2}}\left(
\left\vert 0\right\rangle _{l}+\left\vert 1\right\rangle _{l}\right) $ ($%
l=1,2,3$) [2]. Based on the results (8) and (10) and according to the
expression (6) of the states $|0\rangle $ and $|1\rangle $ and
$U = U_1\otimes \prod_{l=2}^3 U_{1l}$, it is
straightforward to show that for $t=\pi /\chi _{1l}=2\pi /\lambda _{1},$ the
unitary operator $U$ transforms the initial state $|\psi \rangle _{\mathrm{cq%
}}$ of three cqubits as follows
\begin{align}
& \frac{1}{2\sqrt{2}}\left[ \left\vert 0_{1}\right\rangle \left( \left\vert
0_{2}\right\rangle +\left\vert 1_{2}\right\rangle \right) \left( \left\vert
0_{3}\right\rangle +\left\vert 1_{3}\right\rangle \right) \right.  \nonumber
\\
& +\left. \left\vert 1_{1}\right\rangle \left( \left\vert 0_{2}\right\rangle
-\left\vert 1_{2}\right\rangle \right) \left( \left\vert 0_{3}\right\rangle
-\left\vert 1_{3}\right\rangle \right) \right] .
\end{align}

The state (11) can be converted into the following three-cqubit GHZ entangled state
\begin{equation}
\frac{1}{\sqrt{2}}\left( \left\vert 0_{1}\right\rangle \left\vert
0_{2}\right\rangle \left\vert 0_{3}\right\rangle +\left\vert
1_{1}\right\rangle \left\vert 1_{2}\right\rangle \left\vert
1_{3}\right\rangle \right) ,
\end{equation}%
by applying a driving pulse to cavity $l$ to achieve the single-cqubit state
transformation $\frac{1}{\sqrt{2}}\left( \left\vert 0_{l}\right\rangle
+\left\vert 1_{l}\right\rangle \right) \rightarrow \left\vert
0_{l}\right\rangle $ and $\frac{1}{\sqrt{2}}\left( \left\vert
0_{l}\right\rangle -\left\vert 1_{l}\right\rangle \right) \rightarrow
\left\vert 1_{l}\right\rangle $ ($l=2,3$) [2].

The above description shows that the coupler qutrit
remains in the ground state during the entire operation. Therefore,
decoherence from the qutrit is greatly suppressed.

The above condition $\chi _{1l}t=\pi $ and $\lambda _{1}t=2\pi $ turns out
into $\chi _{1l}=\lambda _{1}/2$, resulting in
\begin{equation}
g_{l}=\frac{\left\vert \delta _{l}\right\vert }{\left\vert \delta
_{1}\right\vert +\left\vert \delta _{l}\right\vert }\sqrt{2\Delta
_{1l}\left\vert \delta _{1}\right\vert }.
\end{equation}%
The coupling strength $g_{l}$ can be adjusted by a prior design
of the sample with appropriate capacitance or inductance between the qutrit
and cavity $l$ \cite{s19}.

We now give a brief discussion on the experimental feasibility. Assume that
the single-cqubit operation can be performed within a very short time. Thus,
the decoherence effect is negligible during the single-cqubit operation
and not considered in our numerical simulation for simplicity.

\begin{figure}[tbp]
\begin{center}
\includegraphics[bb=153 561 366 756, width=8.5 cm, clip]{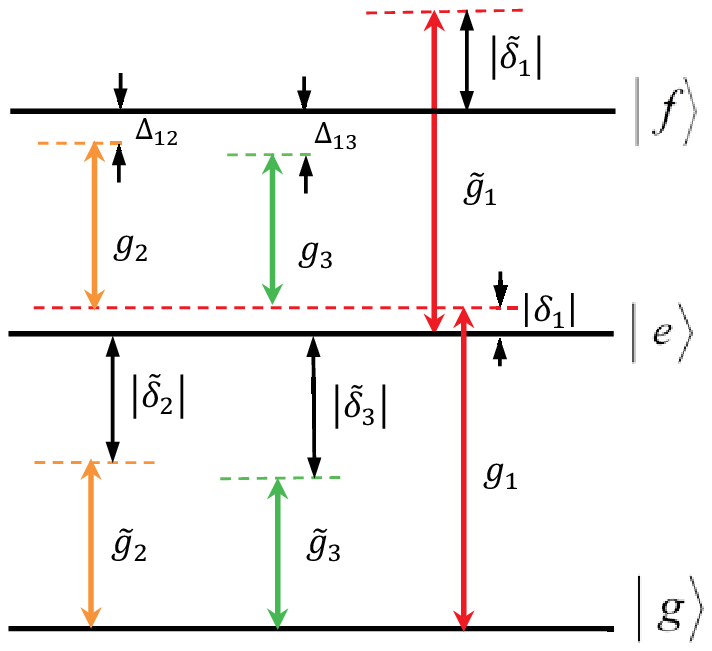} \vspace*{%
-0.08in}
\end{center}
\caption{(Color online) Illustration of the unwanted coupling between cavity
$1$ and the $|e\rangle \leftrightarrow |f\rangle $ transition with coupling
strength $\widetilde{g}_{1}$ and detuning $\left\vert \widetilde{\protect%
\delta }_{1}\right\vert $, as well as the unwanted coupling between cavity $%
l $ and the $|g\rangle \leftrightarrow |e\rangle $ transition with coupling
strength $\widetilde{g}_{l}$ and detuning $\left\vert \widetilde{\protect%
\delta }_{l}\right\vert $ ($l=2,3$). Here, $\left\vert \widetilde{\protect%
\delta }_{1}\right\vert =\protect\omega _{c_{1}}-\protect\omega _{fe}$ and $%
\left\vert \widetilde{\protect\delta }_{l}\right\vert =\protect\omega _{eg}-
\protect\omega _{c_{l}}$ ($l=2,3$).}
\label{fig:3}
\end{figure}

In reality, there exist the inter-cavity crosstalk between cavities,
the unwanted coupling of cavity $1$ with the $
|e\rangle \leftrightarrow |f\rangle $ transition, and the unwanted coupling
of cavities $2$ and $3$ with the $|g\rangle \leftrightarrow |e\rangle $
transition of the qutrit. When these factors are considered, the
Hamiltonian (1) becomes $\widetilde{H}_{\mathrm{I}}=H_{\mathrm{I}%
}+\varepsilon ,$ with
\begin{eqnarray}
\varepsilon &=&\widetilde{g}_{1}(e^{-i\left\vert \widetilde{\delta }%
_{1}\right\vert t}\hat{a}_{1}\sigma _{fe}^{+}+h.c.)  \nonumber \\
&&+\sum\limits_{l=2}^{3}\widetilde{g}_{l}(e^{i\left\vert \widetilde{\delta }%
_{l}\right\vert t}\hat{a}_{l}\sigma _{eg}^{+}+h.c.)  \nonumber \\
&&+\sum\limits_{k\neq l;k,l=2}^{3}g_{kl}(e^{-i\widetilde{\Delta }_{kl}t}\hat{%
a}_{k}\hat{a}_{l}^{+}+h.c.),
\end{eqnarray}%
where the first and second lines describe the unwanted couplings, with
coupling constants $\widetilde{g}_{1}$ and $\widetilde{g}_{l}$ and detunings
$\left\vert \widetilde{\delta }_{1}\right\vert =\omega _{c_{1}}-\omega _{fe}$
and $\left\vert \widetilde{\delta }_{l}\right\vert =\omega _{eg}-\omega
_{c_{l}}$ ($l=2,3$); the last line represents the inter-cavity
crosstalk, with coupling strength $g_{kl}$ and the frequency difference $%
\widetilde{\bigtriangleup }_{kl}=\omega _{c_{k}}-\omega _{c_{l}}$ for
cavities $k$ and $l$.

The dynamics of the lossy system is determined by
\begin{align}
\frac{d\rho }{dt}=& -i[\widetilde{H}_{\mathrm{I}},\rho
]+\sum_{l=1}^{3}\kappa _{l}\mathcal{L}[a_{l}]  \nonumber \\
& +\gamma _{eg}\mathcal{L}[\sigma _{eg}^{-}]+\gamma _{fe}\mathcal{L}[\sigma
_{fe}^{-}]+\gamma _{fg}\mathcal{L}[\sigma _{fg}^{-}]  \nonumber \\
& +\sum\limits_{j=e,f}\{\gamma _{\varphi j}(\sigma _{jj}\rho \sigma
_{jj}-\sigma _{jj}\rho /2-\rho \sigma _{jj}/2)\},
\end{align}%
where $\sigma _{jj}=|j\rangle \langle j|$ $(j=e,f),$ $\mathcal{L}[\xi ]=\xi
\rho \xi ^{\dag }-\xi ^{\dag }\xi \rho /2-\rho \xi ^{\dag }\xi /2$ with $\xi
=a_{l},\sigma _{eg}^{-},\sigma _{fe}^{-},\sigma _{fg}^{-}.$ In addition, $%
\kappa _{l}$ is the decay rate of cavity $l$ $(l=1,2,3),$ $\gamma _{eg}$ is
the energy relaxation rate for the level $|e\rangle $, $\gamma _{fe}(\gamma
_{fg})$ is the energy relaxation rate of the level $|f\rangle $ for the
decay path $|f\rangle \rightarrow |e\rangle $ $(|g\rangle )$, and $%
\gamma _{\varphi j}$ is the dephasing rate of the level $|j\rangle $ of the
qutrit $(j=e,f)$. \newline

The fidelity of the operation is given by $\mathcal{F}=\sqrt{\langle \psi _{%
\mathrm{id}}|\rho |\psi _{\mathrm{id}}\rangle },$ where $|\psi _{\mathrm{id}%
}\rangle $ is the output state of an ideal system without dissipation,
dephasing and crosstalk; while $\rho $ is the final practical density
operator of the system when the operation is performed in a realistic
situation. The ideal output state is $|\psi _{\mathrm{id}}\rangle =|$GHZ$%
\rangle \otimes \left\vert g\right\rangle ,$ where $|$GHZ$\rangle $ is the
GHZ state given in Eq. (12).

\begin{figure}[tbp]
\begin{center}
\includegraphics[width=0.70\linewidth, clip]{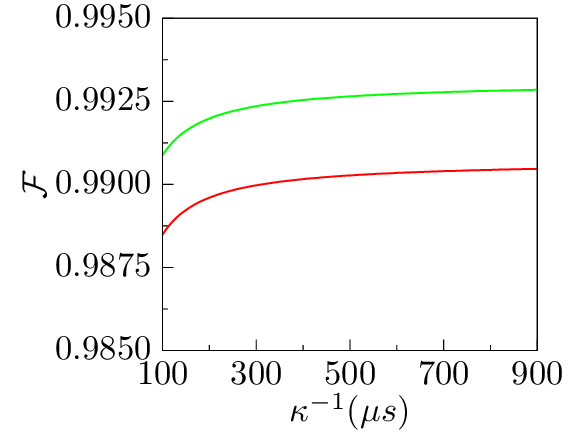}
\vspace*{-0.08in}
\end{center}
\caption{(Color online) Fidelity versus $\protect\kappa ^{-1}$.
The green curve is based on the effective Hamitonian (4) and considering decoherence and the
inter-cavity crosstalk; while the red curve is based on the full Hamiltonian $\widetilde{H}_I$ and considering
decoherence and the inter-cavity crosstalk. From the red curve and the green curve,
it can be seen that the fidelity for the gate performed in a realistic situation is slightly decreased by $\sim 0.25\%$,
when compared to the case of the gate performed based on the effective Hamiltonian (4).
This result implies that the approximations, which we made for obtaining the effective Hamiltonian (4),
are reasonable.}
\label{fig:4}
\end{figure}

For a transmon qutrit, the level spacing anharmonicity $100\sim 500$ MHz was
reported in experiments \cite{s20}. As an example,
consider $\omega _{eg}/2\pi =6.5$ GHz, $\omega _{fe}/2\pi =6.2$ GHz, $\omega
_{c_{1}}/2\pi =7.0$ GHz, $\omega _{c_{2}}/2\pi =5.69$ GHz, and $\omega
_{c_{3}}/2\pi =5.68$ GHz. Thus, $\left\vert \delta _{1}\right\vert
/2\pi =0.5$ GHz, $\left\vert \delta _{2}\right\vert /2\pi =0.51$ GHz, $%
\left\vert \delta _{3}\right\vert /2\pi =0.52$ GHz, $\left\vert \widetilde{%
\delta }_{1}\right\vert /2\pi =0.8$ GHz, $\left\vert \widetilde{\delta }%
_{2}\right\vert /2\pi =0.81$ GHz, $\left\vert \widetilde{\delta }%
_{3}\right\vert /2\pi =0.82$ GHz, $\Delta _{12}/2\pi =0.01$ GHz, $\Delta
_{13}/2\pi =0.02$ GHz, $\widetilde{\Delta }_{12}/2\pi =1.31$ GHz, $%
\widetilde{\Delta }_{23}/2\pi =0.01$ GHz, and $\widetilde{\Delta }%
_{13}/2\pi =1.32$ GHz.

Other parameters used in the numerical simulation are: (i) $\gamma
_{eg}^{-1}=60$ $\mu $s, $\gamma _{fg}^{-1}=150$ $\mu $s \cite{s21}, $\gamma
_{fe}^{-1}=30$ $\mu $s, $\gamma _{\phi e}^{-1}=\gamma _{\phi f}^{-1}=20$ $%
\mu $s, (ii) $g_{1}/2\pi =35$ MHz, (iii) $g_{kl}=0.01g_{\max }$ \cite{s22}, with $%
g_{\max }=\max \{g_{1},g_{2},g_{3}\}$, (iv) $\kappa _{1}=\kappa _{2}=\kappa
_{3}=\kappa $, and (v) $\alpha =0.5$. According to Eq. (13), we have $%
g_{2}/2\pi \sim 50.5$ MHz and $g_{3}/2\pi \sim 72.1$ MHz. For a transmon
qutrit [11], we have $\widetilde{g}_{1}\sim \sqrt{2}g_{1}\sim 2\pi\cdot 49.5$ MHz$,$ $%
\widetilde{g}_{2}\sim g_{2}/\sqrt{2}\sim 2\pi\cdot35.7$ MHz$,$ and $\widetilde{g}%
_{3}\sim g_{3}/\sqrt{2}\sim 2\pi\cdot41.6$ MHz. Here, we consider a rather
conservative case for decoherence time of the transmon qutrit \cite
{s6,s23}. In addition, the coupling constants here are readily available
in experiments \cite{s24}.

By solving the master equation (15), we numerically calculate the fidelity
versus $\kappa ^{-1}$ as depicted in Fig. 4. The red curve is plotted
based on the full Hamiltonian $\widetilde{H}_I$ and by
considering the decoherence and the inter-cavity crosstalk.
From the red curve one can see that when $\kappa ^{-1}\geq 300$ $\mu $%
s, fidelity exceeds $0.9899$. The operation time is $\sim 0.41$ $\mu $s,
much shorter than decoherence time of the qutrit used in the numerical calculation
and the cavity decay times ($100~\mu$s - $900~\mu$s) considered in Fig. 4.
Note that lifetime $\sim 1$ ms of microwave photons was experimentally
demonstrated in a 3D microwave cavity \cite{s6,s25}. For the cavity frequencies
given above and $\kappa ^{-1}=300$ $\mu $s, the cavity quality factors are $%
Q_{1}\sim 1.31\times 10^{7}$ for cavity $1$, $Q_{2}\sim 1.07\times 10^{7}$ for
cavity $2$, and $Q_{3}\sim 1.07\times 10^{7}$ for cavity $3,$ which are
available because a high quality factor $Q=3.5\times 10^{7}$ of a 3D
superconducting cavity has been demonstrated in experiments \cite{s25}.
The analysis here implies that high-fidelity creation of a three-cat-state-qubit GHZ
state is feasible with the present circuit QED technology.

\section*{Funding Information}

This work was supported in part by the NKRDP of China (Grant No.
2016YFA0301802) and the National Natural Science Foundation of China under
Grant Nos. [11074062, 11374083,11774076].


\begin{thebibliography}{99}
\bibitem{s21} For a transmon qutrit, the $|g\rangle $ $\leftrightarrow $ $%
|f\rangle $ transition is much weaker than those of the $|g\rangle $ $%
\leftrightarrow $ $|e\rangle $ and $|e\rangle $ $\leftrightarrow $ $%
|f\rangle $ transitions. Thus, we have $\gamma _{fg}^{-1}\gg \gamma
_{eg}^{-1},\gamma _{fe}^{-1}$.
\end{thebibliography}
\end{document}